\documentclass[manuscript]{aastex}
\newcommand{\HII}{H\,{\sc ii}}

\shorttitle {Massive Stars in NGC 1222}
\shortauthors{Beck et al.}

\begin{document}

\title{The Extraordinary Infrared Spectrum of  NGC 1222 (Mkn 603)}
\author{Sara C. Beck\altaffilmark{1}, Jean L. Turner\altaffilmark{2} \& Jenna Kloosterman\altaffilmark{3}}

\altaffiltext{1}{Department of Physics and Astronomy, Tel Aviv University, Ramat Aviv, Israel email: sara@wise1.tau.ac.il}
\altaffiltext{2}{Department of Physics and Astronomy, UCLA, Los Angeles, CA 90095-1547 email:  turner@astro.ucla.edu}
\altaffiltext{3}{Department of Astronomy, University of California, Berkeley CA 94720}

\begin{abstract}
The infrared spectra of starburst galaxies are dominated 
by the low-excitation lines of [NeII] and [SIII], and the 
stellar populations deduced from these spectra
appear to  lack stars larger than about 35~M$_\odot$. The only exceptions to this result until
now were low metallicity dwarf galaxies.  
We report our analysis of the mid-infrared spectra obtained with IRS on Spitzer of
the starburst galaxy NGC 1222 (Mkn 603). NGC 1222 is a large spheroidal galaxy with 
 a starburst nucleus that is a compact radio and infrared 
 source, and its infrared emission is dominated by the [NeIII] line. 
 This is the first starburst of solar or near-solar metallicity,  known to us, which is dominated by the high-excitation lines and which is a likely host of high mass stars.  We model the emission with several different assumptions as to the spatial distibution of the high- and low-excitation lines and find that the upper mass cutoff in this galaxy is  
 40~M$_\odot$--100~M$_\odot$. 
 
\end{abstract}


\section{Introduction}

Star formation in starburst galaxies is often concentrated in compact, highly obscured regions. It 
is difficult to measure the stellar content of these sources. The stars cannot be studied directly, but only through the nebulae they excite. The proper use of nebular diagnostics has been 
controversial.  In a study of all the galaxies with infrared spectra in the literature as of 2003, \citet{RR04} concluded that the young stellar populations in
galaxies of solar metallicity have relatively low $\rm T_{eff}$ and that their IMFs, if Salpeterian, cannot extend to 100~M$_\odot$.  They suggested that the upper mass cutoffs are $\lesssim$ 40~M$_\odot$.  Dwarf galaxies of much lower than solar metallicity may have higher mass stars; masses as high as 60~M$_\odot$ are needed to explain the spectra of the three such galaxies in 
their sample.  This apparent lack of high mass stars in starbursts is a major puzzle of star formation %
studies. Is it real? Or is the apparent lack due to observational bias or problems in interpretation?%

We report here on the infrared spectra of the unusual S0 galaxy NGC 1222  (Markarian 603;
D=34 Mpc ($\rm h_o/71~$Mpc)$^{-1}$.) This 
galaxy is remarkable because it is only slightly, if at all, less metal-rich than our Galaxy, 
yet has the mid-infrared spectral signature of 
the most  massive, hot stars, including a high ratio of  IRAS 60--100\micron\ flux.   
Our analysis of the galaxy is based on ground-based high-resolution 
maps of the radio and infrared continuum and near-infrared spectra, and on mid-infrared spectra 
obtained with the IRS Spectrometer on the Spitzer Space Telescope. 
We model the mid-infrared emission from this  galaxy as arising in the superposition of 
 a compact and bright nebula that contains the youngest stars and produces the high-excitation ionic lines, and a ``cooler" spectrum region of more extended ($\gtrsim$500 pc) emission.  
The line ratios are compared to the results of STARBURST99 stellar populations and MAPPINGS and
CLOUDY photoionization models.  The models are assumption-rich and the results are only limits.
However, even with these
caveats we can say that this galaxy has the highest upper mass cutoff yet known for a galaxy  
that is not a metal-poor dwarf.  

In the next section we describe the observations and how we model the galaxy, in the Analysis section 
the STARBURST99 and photoionization grids, and in the Discussion, our conclusions.  

\section{Observations}
\subsection{Previous Work}

\citet{PB93} obtained optical spectra and images of the Markarian galaxy,
NGC~1222. The galaxy has the optical
appearance of a triple system. 
They found that the main source has a starburst nucleus of high surface brightness, superimposed on a fainter elliptical component of approximately 13 kpc by 9 kpc extent to the 25 mag arcsec$^{-2}$
isophote.    
Within 2\arcmin\ (20 kpc) of 
the nucleus are two compact sources of high surface brightness, C1 and C2, which they argue are dwarf galaxies interacting with the main galaxy.  In spite of the proximity of the compact components, the
overall isophotal morphology, even at large galactocentric radii, remains fairly symmetric. 
The optical spectra 
show that all three sources are photo-ionized, rather than shock excited, and that the main source
has the characteristic spectrum of a starburst.    
\citet{PB93} found elemental abundances
of O, Ne and N to be $\rm 12+log[x/H] ~8.57\pm0.09, 8.53\pm0.026$ and $7.35\pm0.17$ respectively.  
\citet{CDD01} 
 obtained K-band spectra of NGC 1222 and report the Br$\,\gamma$ flux of the galaxy in 
2 apertures: $\rm 23.5\pm0.8\times10^{-15}erg\,s^{-1}~cm^{-2}$ 
in a large beam of 3\arcsec$\times $9\arcsec\ and
$16.9\pm 0.4$ in a 3\arcsec $ \times$ 3\arcsec\ beam, also consistent with a bright, compact starburst
with an additional extended ionized component. H$\alpha$ + [NII] images of NGC~1222 show that
the ionized gas is brightest at the locations of the compact continuum sources \citep{AHM90}.

NGC 1222 is a strong infrared source and is in the Revised Bright Galaxy Sample
\citep{2003AJ....126.1607S}.  
Its IRAS fluxes are 0.59, 2.28, 13.06  and 15.41 Jy at 12, 25, 60 and 100\micron\ respectively; it has the 
high ratio of 60/100\micron\ flux  typical of active starbursts.  
\citet{E96} searched for
CO in NGC 1222 and did not detect any, so the molecular gas mass is not directly detected. 
However, dust emission in NGC 1222 has been detected at $850\mu$m by 
\citet{2000MNRAS.315..115D} using SCUBA   
and it has been mapped in  $HI$ with the VLA by 
\citet{2004MNRAS.351..362T}  
The logs of the atomic gas and dust masses are 9.38 and 6.66; NGC 1222 is at the low end of the sample for both quantities.   

\subsection{Radio Continuum Maps}

The spatial distribution of the starburst is a key element of the interpretation of
the mid-infrared spectra, and NRAO VLA\footnote{The National Radio Astronomy
Observatory is a facility of the National Science Foundation, operated under
cooperative agreement by Associated Universities, Inc.}
 maps allow us to map the structure of the
starburst. We show radio maps at 6~cm (C band) and 1.3~cm (K band) 
obtained from the VLA archives in Figures 1a and 1b. The 
1.3~cm image
 is shown superimposed on the 
 6~cm in Figure 1c. Details of  the observations are in Table 1. 
 The beam sizes of the observations differ by about a factor of two; 
 for the overlay in Figure 1c, the maps have been convolved to the same
 1.75\arcsec $\times$ 1.2\arcsec\ beam. The VLA maps are sensitive to 
 structures less than $\sim$15\arcsec\ in size. 

In the 6~cm map, there is a secondary source about 10\arcsec\ east of the main source, and there is a $3\sigma$ feature in K-band at the same location. Comparing the radio 
data to the
only image of \citet{PB93}, we associate the second radio source
with their C1 object, which they identify as
a separate interacting system.  We see no radio 
counterpart to their C2 to the southwest of the main source, although it is a bright
H$\alpha$ + [NII] source \citep{AHM90}.    \citet{C90} find a size of 11\arcsec\ $\times 8$ \arcsec\ 
at 20 cm and gives a total 20 cm flux of 55 mJy in 18\arcsec\ and 39.7 in 5\arcsec . 
The C1 source may appear
as an extension in their map. The position of our radio source agrees with that of \citet{SW86}
20~cm map 
and is about 10\arcsec\ off the quoted optical source position, but that position is quoted with large 
($\pm5$\arcsec) uncertainties.

The radio emission at 1.3~cm 
comes almost entirely from a bright source about 4\arcsec\
(620 pc) in diameter; the 1.3~cm 
emission has an unusual shape which could be an incomplete ring, or two bright peaks not
completely separated.  The 6~cm  
emission has the same bright source with structure consistent
with the 1.3~cm 
shape; the beam size at 6~cm  
is almost twice that at 1.3~cm  
so the same detail cannot be seen. 
The 6~cm map  
 also shows extended emission, mostly in the east-west direction. 
The radio maps are consistent with the 
$11.7\mu$m image in the Keck/LWS images of Turner, Gorjian \& Beck 2007, which 
also shows a bright, concentrated source extended EW.  
Neither C1 nor C2 fall in the field of the LWS observations.

Comparison of the VLA maps at 6 and 1.3~cm, and with single dish fluxes,
reveals much about the spatial distribution of the radio emission and the starburst.
 The galaxy looks very different in the two images, so there are clearly
variations in spectral index and sources of the radio emission.  
The expected sources of radio emission in starburst galaxies are nonthermal
synchrotron emission, with spectral index, $\alpha \sim -0.7$ to $-0.8$, and
thermal free-free emission from HII regions, with $\alpha \sim -0.1$ ($S\propto \nu^\alpha$). 
The 1.3~cm  source has flux, 
found using the AIPS program TVSTAT, of $S_{1.3\,cm}=14 \pm 2$~mJy in 28 square arcseconds.
 The central source at 6~cm,  
 excluding the extended emission components, has $S_{6\,cm}=14 \pm 4$~mJy in the same region. 
The central source is thus a bright 1.3~cm emitter, roughly as bright at 1.3~cm  
as at 6~cm, and the radio emission in the central region is mostly if not entirely thermal.  
Single dish fluxes measured by  \citet{marx94} with the Effelsberg 100~m telescope are
 30.7 $\pm 3$ mJy at 6~cm and 14.8 mJy $\pm 3$ at 
 2.8 cm; the 100~m beamsizes are 145\arcsec\ at 6~cm and 69\arcsec\ at 2.8~cm. 
Comparison of the total mapped flux in our 6~cm VLA map to the Effelberg fluxes 
indicates that there is a significant extended component to the 6~cm emission
that is not detected in our VLA maps. This
extended  component appears to be dominated by nonthermal 
synchrotron emission, based on the single dish spectral index of -0.9. 
By contrast, the 1.3~cm VLA flux is essentially equal to the 2.8~cm Effelsberg flux. Since
optically thin thermal free-free emission from $H_{II}$ regions has a nearly flat spectrum, 
with a $\nu^{-0.1}$ spectrum,
 then the VLA map must detect nearly all of the 1.3~cm emission in this galaxy,
 and the source must be largely free-free emission.  Although a strong thermal radio 
 source, this central source is not the brightest source in
 the image of H$\alpha$ + [NII] image of \citet{AHM90}, probably because of extinction
 ($\S 3.2$).
 
 Thus the high frequency radio fluxes indicate that the 1.3~cm VLA map of NGC~1222 is
 mostly thermal free-free emission
 from HII regions, and that this emission is confined to the source 
 shown in the map of Figure 1b. In the VLA maps, the 1.3~cm emission 
in NGC 1222 appears even stronger relative to the 6~cm emission than the slight negative  
spectral decrement of thermal emission would require. 
This source may be one of those, common in extreme starbursts, where the radio spectrum rises instead of falling at high frequencies because lower frequencies have significant optical depth 
\citep{THB98,BTK00,J01}.

The inner region of NGC 1222 contains a large population of young ionizing stars. If we 
assume the 1.3 cm emission in the VLA maps  is entirely thermal free-free emission, the 
 observed flux implies $\rm N_{Lyc}$ of $\rm1.45\times10^{54}~s^{-1}$ for the
central 3\arcsec.  This is the ionization equivalent of  $1.45\times10^5$ standard O7 stars.

\subsection{Spitzer IRS Spectra}

NGC 1222 was observed with the IRS on Spitzer in staring mode as part of the IRS Standard Spectra 
program, Program 14, for which J. Houck was Principal Investigator .  The middle-infrared spectral diagnostics we need for this study fall in the bandpass of the
Short-Hi module, which covers the wavelength range 9.9-19.6\micron\ with spectral resolution 
of $\approx 600$. The NGC 1222 data were obtained in two positions separated by about 3\arcsec\ 
roughly parallel to the slit.  Their full Spitzer identifiers are in the Table;  here they will be called Spectrum 1 and Spectrum 2.   Spectrum 1 is shown in Figure 2. The line fluxes obtained for the diagnostic lines, for
each spectrum,  are shown in Table 2.  
We extracted the spectra from the post-BCD data provided by the Spitzer pipeline.    IRS data requires different treatment for extended and point sources. 
Since NGC 1222 is neither a point source nor 
a flat and smooth extended source, we follow the SINGS group \citep{seth06},
 who use the extended source option in the SPICE package to obtain line over continuum fluxes in clumpy galaxies. They have shown that  where  the flux calibration can be constrained by
other data, the extended source calibration is usually accurate to within 20\%. The uncertainty in the fluxes and the line ratios is dominated by the calibration difficulties: since the correction for spatial structure depends strongly on the wavelength, both the relative and
the absolute calibration are affected.  We assume 20\% as the working uncertainty for all the lines.  

The nomimal position of Spectrum 2 is 03$^h$ 08$^m$56.7$^s$, 
$-$2\arcdeg 57\arcmin 18\arcsec\ and of Spectrum 1 
03$^h$ 08$^m$ 56.9$^s$, $-$2\arcdeg\ 57\arcmin\ 20\arcsec; the former 
agrees better with the radio position. However, the beam size in the IRS is large, on these scales, and also depends on wavelength: the beam is around $3\arcsec$ at $10\mu$m and $6\arcsec$ at $20\mu$m, so the positions actually overlap for most of the wavelength range.    Spectrum 2 has somewhat stronger lines than Spectrum 1 but the difference is less than the calibration uncertainty in all cases, and the line ratios on which the 
analysis in the next section depends differ by less than $10\%$ between the two observations. We have
therefore taken the average of the line ratios for each observation and work with that number for
simplicity. 
 
\section{Analysis}

\subsection{Spatial Stucture} 

The ground-based radio, K-band  and $11.7\mu$m continuum data give a consistent picture of  NGC 1222: the young stars are
concentrated into a central source no larger than 3\arcsec\ diameter, $\sim 470$pc,  
which produces most of the 
infrared and radio emission in the galaxy.  This is the source called the starburst nucleus by 
\citet{CDD01} and by \citet{PB93}.
We do not know if it is the kinematic nucleus of the galaxy, as there is little dynamic information, but it is clearly the dominant region of star formation and the actual starburst. (Note that the Spitzer Short-Hi observations  do not cover the
optical and 6~cm 
source C1, and we cannot judge whether it is also an RISN or some
other kind of emission region.) There are more young stars in a less dense region 2-3 times as large as the compact source. The phenomenon of 
a very concentrated group of young stars with strong radio and dominant infrared emission is a common
one in starburst galaxies; we have called such sources in other galaxies RISN 
\citep[Radio-Infrared Super-Nebulae][]{B02}.

How are we to analyse the infrared spectra in light of  the observed spatial structure of the galaxy? The 
suite of infrared diagnostic lines can be for convenience thought of as the high-excitation lines, [NeIII] and [SIV], and the low-excitation lines [NeII] and [SIII].  (The  [ArII] and [ArIII] fine-structure lines
at 6.99 and 8.99~\micron\  are also useful diagnostics, but do not fall in the wavelength region
measured by the Spitzer high resolution modules, and cannot be seen clearly in the low-resolution 
results). The homo- and hetero-nuclear line ratios give a measure of the relative strength in the total radiation field of the hard photons that create the ions of $Ne^{++}$ and $S^{+3}$  and the softer photons responsible for the other ions.  From this measure
of the ionization it is possible to work backwards to stellar populations that could produce that
ionizing field. This is an active area of research and there are several programs and approaches, comprehensively 
reviewed in \citet{RR04}. 

For the line ratios to be meaningful they must compare lines from the same source. 
The galaxy appears in both the radio and optical maps to have a compact main source embedded in more diffuse emission.   The IRS slit is
large enough to encompass 
 most of the total emission, compact and diffuse together (Figure 1).  
If the high- and low-excitation 
lines are distributed differently between the compact and diffuse components, the
global line ratios will
be subject to misinterpretation.
This phenomenon has been observed in other galaxies, 
particularly those that are forming super star clusters, such as NGC 5253 \citep{C99}.   
It is also what we should expect, from the nature of the IMF. 

In any IMF the most massive stars will be only a small fraction of the total by number.  To have enough very massive
stars that they significantly affect the spectrum, the total number of stars must be very large.  
So the most massive stars will be found in NGC 1222, if at all,  
in the compact source, which as shown above  must contain tens of 
thousands of O stars.   A star must be larger than 
40-45 $M_\odot$ to excite as much [NeIII] as [NeII] emission.   [NeII], in contrast,  is excited by the 
more common late-O and early B stars: even a B0 star of only 20$M_\odot$ 
 can produce strong [NeII] emission.   
 These lower mass stars are a much larger fraction of the total by number and will
 be common even in the sparser stellar population of the diffuse source. 
 
So we argue that almost  all the [NeIII]  should be 
assigned to the compact 3\arcsec\ source visible in the VLA 1.3~cm image. 
But what about the [NeII] and the sulfur lines? 
What fraction of the line strengths should be attributed to the 
compact source and what to the extended? We note that \citet{CDD01}
found an excess of $\rm 6.6 \times 10^{-15} erg\,s^{-1}\, cm^{-2}$  of
 Br$\,\gamma$ emission in their large beam over that in their small. 
 Their small beam being comparable in size to the compact source, 
 and their larger beam to the entire Spitzer slit, we
take this extended component of emission and estimate the [NeII] flux that should be produced
by that quantity of ionized gas.  We use the neon abundance found by \citet{PB93} 
of $3.34\times10^{-4}$ 
relative to H and assume that all the neon is
singly ionized; the latter is the default case in normal {\HII} regions.  With these assumptions
the 
extended 
 component is expected to produce a [NeII] line of 1.8 Jy strength. Which equals,  within the 
calculational uncertainties, the entire [NeII] line flux seen by Spitzer!  
This argues that the low-excitation line of [NeII] is 
almost entirely produced in the extended emission and that
the contribution of the compact source is negligible. 

  So NGC 1222 appears to 
resemble NGC 5253 \citep{C99}, where all the [NeIII] flux seen in the much larger beam of ISO 
proved to be emitted by a single compact source, and all the [NeII] from an extended emission region. 
How will the other lines track?  We don't know enough of the structure of
the ionization field to treat this fully, but can get a crude and approximate answer from the ionization 
potentials of the sulfur ions.  $S^{++}$ 
can only exist in the $Ne^+$ zone, but $S^{+3}$ can coexist with $Ne^+$ as well as with $Ne^{++}$. So the extended diffuse emission could produce
some part of the [SIV] as well as essentially all of the [SIII].  In NGC 5253, this was not the case, and all the [SIV]  was produced by the compact source, but in II Zw 40 (another low-metallicity dwarf) as much
as 30\% of the SIV may come from outside the compact source \citep{MH06}. 

Since we do not know the spatial distribution of the mid-infrared lines, we therefore work with the line ratios we would derive for the following three cases. These cases should cover all possibilities for
the distribution of the mid-infrared lines in NGC 1222.

{\it Model 1: } no spatial separation; the lines are co-extensive in origin.  %
We know from the $Br\gamma$ results  and all the
arguments above that this model is unrealistic. 
Nevertheless, we present it as an extreme lower limit to the stellar temperatures
and masses. 

{\it Model 2:} the most stratified model; the high and low-excitation lines are essentially disjoint. %
In this case NGC 1222 would resemble
NGC 5253, where all the low-excitation lines were from the diffuse component. For the calculations we assume that 
$90\%$ of the [NeIIII] and [SIV] are from the  compact source and $90\%$ of the [NeII] and SIII] from the diffuse.  We 
chose $90\%$  because the remainder it leaves is consistent with the weakest Spitzer lines and the  calculated [NeII] flux.  This will give the hottest and most massive stars. It is also, we think, the most realistic model. 

{\it Model 3:} an intermediate case, found by applying the solar neon abundance rather than %
the higher value found
by \citet{PB93} to the gas producing the extended $Br\gamma$ emission. 
 This gives  0.96 Jy of [NeII] in the extended emission, for a model in which 60\% of the low excitation lines [SIII] and [NeII] are in the extended source and 40\% in the compact. In this
model as in model 2,  $90\%$  of the high-excitation lines are from the compact source; the distributions of [NeII] and [SIII] 
are changed.  

    These physical models are simple but they are consistent with 
everything we know about the source and about similar star formation regions and they will show how the stellar types deduced depend on the spatial distribution. Line ratios for the models are given in Tables 3 and 4.

\subsection{Extinction Effects} 
In the above discussion we used the observed line strengths without correction for extinction.  
Based on optical spectra, \citet{PB93}    found the maximum 
reddening to be 0.4 mag, for a visual extinction  of 1.2 mag. 
  The $Br\gamma/$radio ratio tells 
a different story; the $Br\gamma$ fluxes are considerably lower than the 
value predicted from the radio continuum flux, which for $T_e=7500$~K 
and no extinction at $Br\gamma$ 
would  give $\rm 1.6\times10^{-13}erg\,s^{-1}\, cm^{-2}$ in 3\arcsec$\times$3\arcsec. 
The discrepancy with the observed gives 2.4 mag of extinction at 
B$\,\gamma$ and 19 magnitudes $A_v$.  The different extinctions derived from different markers are common in star formation regions, where
the optical emission region and the infrared sources can be physically distinct, and where much 
of the extinction is produced within the HII regions themselves.

An obscuration of 19 magnitudes is typical of very dense compact $H_{II}$ regions. If we accept 
that value for the region producing the infrared lines in NGC 1222 we will have to consider the
effect on our line ratios.  We follow \citet{G02} and take $A_{SIV}=0.7 A_k$, $A_{NeII}=0.28 A_k$, 
$A_{NeIII}=0.22 A_k$, and $A_{SIII}=0.32 A_k$. The extinctions at the  [NeII], [NeIII], and [SIII] lines are
so similar that even with this high figure for the total extinction the line ratios will change by only about
20\%. The uncertainty introduced by the extinction is thus less that that due to the spatial structure, the
metal content, and the instrument calibration.  The problem is more severe for [SIV], where the
line fluxes would be increased by a factor of 4.8.  Correcting the [SIV]/[SIII] and [SIV]/[NeII] ratios 
for this extinction would increase them  by factors of 2.3 and 2.5, respectively.   This extinction factor would 
move the line ratios calculated for model 1 close to those of model 3, and those of model 3 close to model 2. 

The distribution of obscuring matter in NGC 1222 is probably not uniform, just
as the emission is not uniform.  The extinction may be higher towards the compact source and 
lower in the extended emission region.  Until the galaxy is mapped with high resolution in the
infrared, the above results are only approximate. 

\subsection{Photoionization Models}

Deducing the stellar population from the observed infrared lines is a complicated process. The line ratios are compared to those predicted by models, and the inputs into the models, 
which include the time dependance 
of the star formation process (i.e., is it an instantaneous burst or continuous), the mass function and mass
limits of the burst, the stellar atmospheres, and the metal content of the stars and of the nebula.  Much
work has gone into the establishment and testing of the photoionization models and improvements
continue, especially on the stellar atmospheres, the inclusion of stellar winds, and the inclusion  of the Wolf-Rayet stage.  The models are constantly changing and improving. 

We use the STARBURST99 software and data package \citep{L99} to model the ionizing 
spectra produced by different candidate stellar populations, and the MAPPINGS code \citep{K07} to calculate
the output spectrum.  This method gives results generally in good agreement \citep{RR04} 
with output spectra calculated by  CLOUDY \citep{F06} from ionizing spectra generated by
STARBURST99.  The models are most limited, or least realistic, in their geometrical 
simplicity---real star formation regions are not likely to be spherical, symmetric or uniformly filled.   In NGC 1222, as in almost all galaxies,  we know that there is complex structure
but it can't be observed directly. 
So we use the default geometrical assumptions in the photoionization code of sphericity, uniform filling, and isobaric 
density structure with $P/k=10^5$ and mean temperature $10^4$.  The ionization parameter is set by 
the STARBURST99 output luminosity and the density. The models compare bursts of $10^6$~M$_{\odot}$  with a Kroupa IMF with power-law exponents  1.3 for 0.1  to 0.5~M$_\odot$  and 2.3 for higher masses. We assume Pauldrach/Hillier stellar atmospheres, 
Geneva evolution tracks with high mass-loss rates, 
and dust of
standard depletions. 
 We vary the parameters 1) upper mass limit of the IMF, which may be refered to as the mass and 2) the
age.  We also ran models for different values of metal content. 
 
The infrared  diagnostic lines are very sensitive to the metal content of the source.  Starbursts with 
lower metal content will have hotter nebulae than others of the same mass parameters and ages but
with more metals.  This is partly because stars with fewer metals, and thus less line-blanketing, will have hotter spectra than metal rich stars of the same mass, and partly because the ionized nebula cool by emitting metal lines.  Until now, the starbursts  in which [NeIII] is clearly stronger than [NeII] have been 
low metal-content blue dwarf galaxies: NGC 5253, II Zw 40, and NGC 55. The metal content of NGC 
1222, in contrast, is close to Galactic: \citet{PB93} describe it as ``only slightly, if at all
deficient", with [O/H] $70\%$ of solar, [Ne/H] almost twice solar, and only [N/H], at 35\% of solar, really deficient.  Other elements were not measured.  STARBURST99 offers stellar atmospheres in a range of metallicities, but the only realistic options for this galaxy were 0.4 solar, solar or twice solar. We used  solar value as it is closest to the reported. The metal content of the 
nebula can be adjusted in the codes and  we used gas with the stated abundances of O, Ne and N and the solar abundance of all others. 

A weakness of the models is that the stellar atmosphere abundance cannot be fine-tuned, and if the stars in NGC 1222  have metal content significantly lower than solar, 
they will produce more high-excitation gas  than predicted and the stellar types derived from the models
may be misleadingly high.  We tested the possible impact of this by running several models with stellar atmospheres of 0.4 solar metallicity. If the nebular gas abundance patterns were the same, the two stellar atmosphere models produced line ratios within 15\% of each other.  So unless the stellar metal abundances are radically different from the estimated, this 
will not strongly affect the results. 
 
\section{The Most Massive Stars in the Starburst}
\subsection{The Compact Source}

In Figure 3 we show the inferred line ratios we would observe
for the three models of spatial distribution, overplotted on simulation results. 
For Model 1, which treats the whole
galaxy as one uniform source of all mid-infrared lines, 
and will give the coolest, least massive stars, the [NeIII]/[NeII] ratio can be fit by age less than 
$~2.2$~Myr and an upper mass limit of 35~M$_{\odot}$, or alternatively,
age 2.2--3.2~Myr and any mass above 40~M$_{\odot}$.  In Model 2, which we favor because of the radio morphology, 
the extreme concentration model in which nearly all the low excitation emission comes 
from extended ($>$500 pc) gas,
only an upper limit of $\gtrsim 50-70~M_{\odot}$ stars and
age less than 2 Myr can match the results. For 
Model 3, the intermediate case in which 40\% of the low excitation [SIII] and [NeII] flux is emitted by the nuclear region,
 the neon line ratio gives an upper mass limit around 40M$_{\odot}$ for  age less than 2 Myr or 
 up to 100~M$_{\odot}$ if in the 2--3 Myr range where the ionization is coolest.

[NeIII]/[NeII] has two clear advantages over the other line ratios we can form with the 
Spitzer data: it is homonuclear, and the two lines have very similar, very low, extinction (the line
ratio does not have to be corrected for extinction at the current level of analysis).  The [SIV] line is 
very close to the silicate absorption feature and has the highest extinction of the useful infrared lines, and the 
sulfur line ratio, since it covers the greatest wavelength range, is also the most susceptible to the
wavelength dependance of the extended source calibration.  
With these warnings, the [SIV]/[SIII] line ratio without extinction correction is best fit by, in 
Model 1: masses less than 30~M$_{\odot}$ and ages less than 3~Myr; 
Model 2: masses greater than 50~M$_{\odot}$ and ages less than 2.5~Myr, and Model 3: masses of 30~M$_{\odot}$ and ages less than 3~Myr. 
Extinction corrections move these results towards higher masses; the corrections in 
section 3.2 will give mass limits of about 30, 40, and $>$70~M$_{\odot}$ for the three spatial 
models. 

The [SIV]/[NeII] line ratio can be affected both by the significant extinction and by abundance. The simulations
show that  model 1 can be fit with 30--40~M$_{\odot}$ stars less than 2~Myr old, 
 model~2 by 
$>$50~M$_{\odot}$, and model 3 by a mass $>$40~M$_{\odot}$. The extinction correction results moves the best fit to 40 and 70~M$_{\odot}$ for
models 1 and 3 and more than 100~M$_{\odot}$ for model~2.

Note that in many cases the line ratios can agree with either a very early age, or a time later than
about 3.5~Myr when stars have entered the Wolf-Rayet phase, which produces a hard ionization field. We did not 
consider the second stage because the Wolf-Rayet feature has not been seen in NGC~1222.   But the  high extinction and
the paucity of deep optical spectra in this galaxy mean that Wolf-Rayet stars have not been strictly ruled out, and the 
possibility should be kept in mind. 

\subsection{The Diffuse Emission}
The [NeIII]/[NeII] ratio in the diffuse region, although based on assumptions, is plausible because 
the [NeIII] is almost certain to be strongly concentrated and because the [NeII] flux is derived 
in a fairly straightforward way.  The [SIII] cannot be so easily calculated because the abundance of sulfur
 is not known.  With these considerations, the line ratios agree with a relatively low upper
 mass limit of 30 M$_\odot$, and age less than 4~Myr.  In this regime, Models 2 and 3 fit the 
 same photoionization grids equally well.  
 
\section{Discussion }
Deducing the stellar population of a star cluster from the nebular line it excites is never straightforward; we must find the  best values of mass and age with only three line ratios observed.  
In NGC~1222
it is even more complex, because of the large slit on Spitzer which combined flux from the compact
source and from the extended emission, so that we must dis-entangle the spatial distribution from the
observed line ratios.  We have presented many models and must now consider which are the most
physical and probable.  

That the compact source in NGC~1222 must contain most of the [NeIII] and [SIV], and that the
extended Br$\,\gamma$ emission must be associated with some [NeII] and [SIII], is certain. 
Reality is thus 
somewhere between our Models~2 and 3 (the most concentrated and intermediate cases).
Among the line ratios, we 
give the highest weight to [NeIII]/[NeII] as it is homonuclear and
not strongly affected by extinction.  (The [SIV]/[SIII] ratio compares two lines whose critical densities are
significantly different, $3.7\times10^4$ for [SIV] and $9\times10^3$ for [SIII],  both within the range of densities expected for compact star formation regions, 
in addition to the considerable extinction correction.) So the best-fitting models are those with upper mass limit  between 40~M$_\odot$ and 100~M$_{\odot}$  and age
less than 2~Myr.   These results cannot 
be better refined until high resolution observations of the infrared lines, which would directly measure
the contributions of the compact and diffuse component, can be carried out.  

\subsection{Comparision to Other Simulation Results}
We compare our model results to the CLOUDY simulations of \citet{RR04}.   They used the same STARBURST99 input spectra but CLOUDY instead of the Mappings photoionization code;  
in Figure 11 of their paper it may be seen that for high masses and solar metallicity they agree 
very well.   We show in Figure 4 the line ratios as a function of time for CLOUDY simulations and
solar metallicity, taken from that paper.   The [NeIII]/[NeII] line ratio of NGC~1222 is so high that 
even the unrealistic, least concentrated Model 1 would require an upper mass limit 
of $\sim 50~M_{\odot}$, and for the other models even 
100~M$_{\odot}$ would not provide enough ionization. The [SIV]/[SIII]
and [SIV]/[NeII] ratios behave similarly.   

The MAPPINGS models of this paper and the CLOUDY models of \citet{RR04} used different parameters
and structures for the ionized regions, but it is apparent that the nebular metallicity is the driving factor that makes the results so different.  If the metallicity of NGC 1222 were  0.2 solar, it would be like NGC 5253, which has similar line ratios, and which  (Figure 4 of Rigby \& Rieke)  can be fit with M$_{upper}>40$M$_{\odot}$ .  
In a solar metallicity system, the line ratios of NGC 1222 can only be fit by truly extreme stellar populations.  In the moderately sub-solar metal models which are most appropriate for the galaxy, the stellar population deduced is unusual but not extreme.  
 \section{Conclusions}
NGC 1222 is a remarkable galaxy.  
While there are uncertainties in the models we used for analyzing
ionized gas emission, 
the results are clear and not model-dependent:  
NGC 1222 has  the highest excitation nebular spectrum and the highest deduced upper 
mass limit for the starburst stellar mass function of any
galaxy yet studied that is not a metal-poor blue dwarf.  NGC 1222 has about the same 
infrared diagnostic line ratios as the iconic metal-poor dwarf NGC 5253, but NGC 1222 has near-solar
metallicity.  It is unique among currently studied infrared galaxies.

The infrared-line ratios of NGC 1222 would be extreme, for a galaxy of near-solar metallicity,  even
without the extra factor added by our spatial modeling. With the models, the line ratios inferred are unique. That the inferred line ratios do resemble those observed in low-metal dwarfs suggests that perhaps the infrared emission
is generated in a low-metallicity region.  It would explain the results if, for example,  NGC 1222 is actually a low-metal content dwarf that has merged with a system of higher metal content, and  if
\citet{PB93}  measured their optical spectra in an area heavily contaminated with metals from the second system.  The fact that there is a bright companion within 2 kpc of the central source of NGC 1222 
indicates that the galaxy is currently experiencing a significant interaction. 
It would also would be consistent with another anomaly of NGC 1222: its low molecular gas content  
\citep[attempts to measure the CO have only set upper limits:][]{C92,E96}.  
Starburst galaxies are usually rich in molecular gas unless they are metal-poor dwarfs.   But NGC 1222 cannot be a mis-classified dwarf galaxy, because it is too big and too bright.   It has roughly half the infrared flux of NGC 5253, yet is 10 times further away. Its total luminosity puts it well outside the range of a dwarf galaxy.

 Higher spatial resolution measurements of the 
mid-infrared lines, specifically [NeII] and [SIV] which can be observed 
from the ground with sub-arcsecond resolution, would permit us to see directly the compact and extended 
components, and to determine how much extinction affects each. Higher resolution maps
in the radio would find the spatial structure. The source cannot be a single cluster: at  1.3 cm its
 diameter is over 500 pc and its ionization is equivalent to  $1.4\times10^5$ O7 stars, 
 twenty times that of the ``supernebula" in NGC 5253.  
 If the ionization is produced by stars with a Salpeter IMF and  a lower mass cutoff of 1 M$_\odot$ the total mass in stars is $2-3 \times10^7$ M$_\odot$. 
 This exceeds by a large factor the most massive star clusters known.  The high 1.3~cm 
flux raises the possibility that there are very compact, dense, rising-spectrum sources 
in NGC 1222. This would put it in the class of the most extreme and youngest starbursts, sources
that are currently not well understood.   

Why is this galaxy so different from most solar metallicity galaxies?  The answer may be in the 
two companions which it is either absorbing or otherwise interacting with.  A merger with one companion usually results in 
an intense burst of star formation. Perhaps  in a double merger the process becomes especially extreme. 

\section{Acknowledgements}

This research has also made use of the NASA/IPAC Extragalactic Database (NED) and the NASA/ IPAC Infrared Science Archive (IRSA) which are operated by the Jet Propulsion Laboratory, California Institute of Technology, under contract with the National Aeronautics and Space Administration. We are
grateful to Dr. V. Gorjian for IRS advice and to Dr. J. Rigby for use of the CLOUDY figures and helpful discussions. 

{\it Facilities:} \facility{VLA}, \facility{Spitzer}.

\begin{figure}
\epsscale{1.0}
\plotone{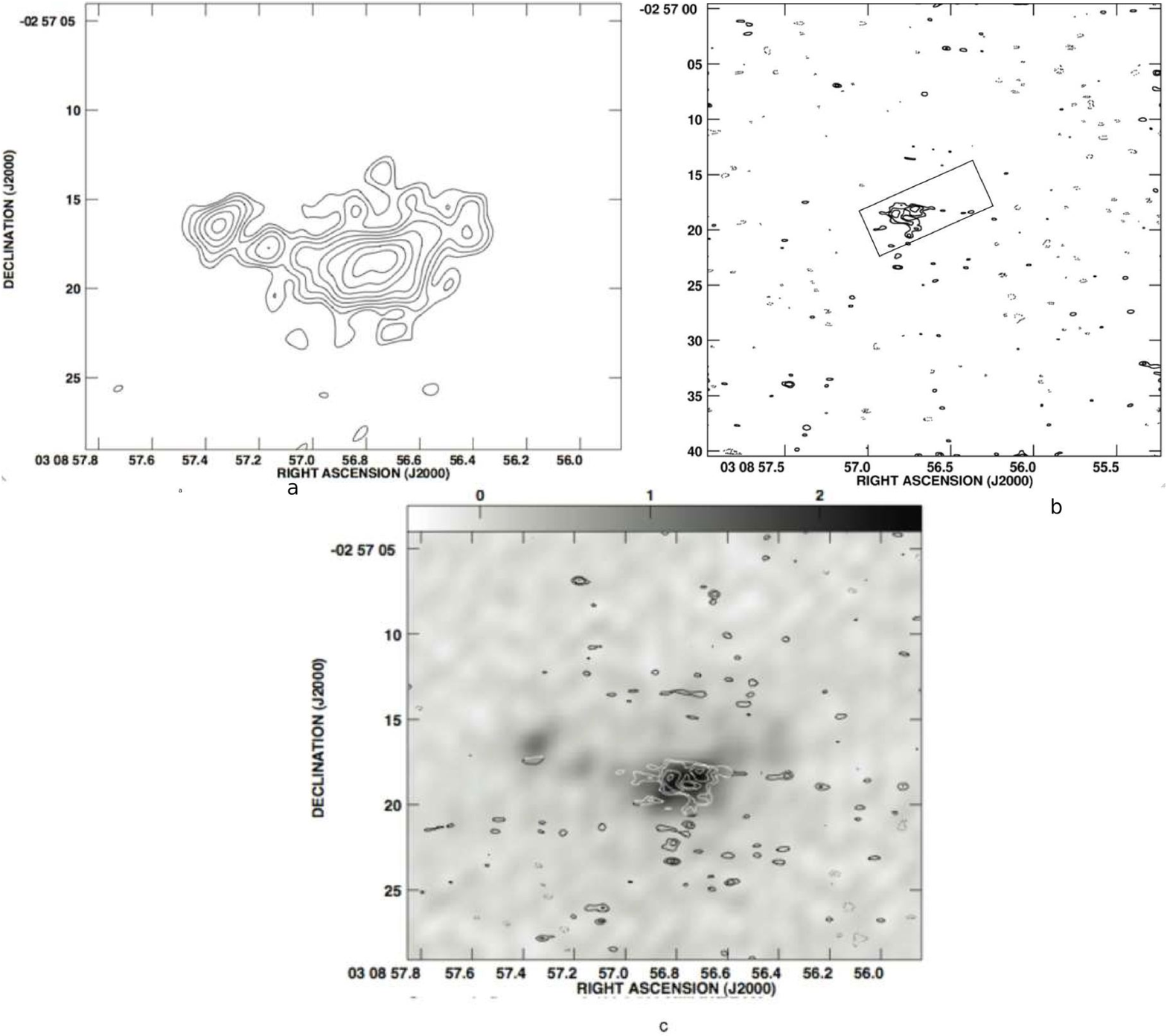}
\caption{Radio maps of NGC 1222. 
{\it top left:}\  6~cm image. Contour levels  are $\pm 2^{n/2} \times$ 0.18~mJy/beam. 
{\it top right:}\ 1.3~cm image. Contour levels are $\pm 2^{n/2} \times$ 0.35~mJy/beam. 
The Spitzer IRS slit is superimposed. 
{\it bottom:}\ Contours of 1.3 cm emission on grey scale of 6-cm.  Grey-scale is in mJy/beam,
shown at top.  Contour levels are $\pm 2^{n/2} \times$ 0.15~mJy/beam.}
\label{radio}
\end{figure}

\begin{figure}

\epsscale{1.0}
\plotone{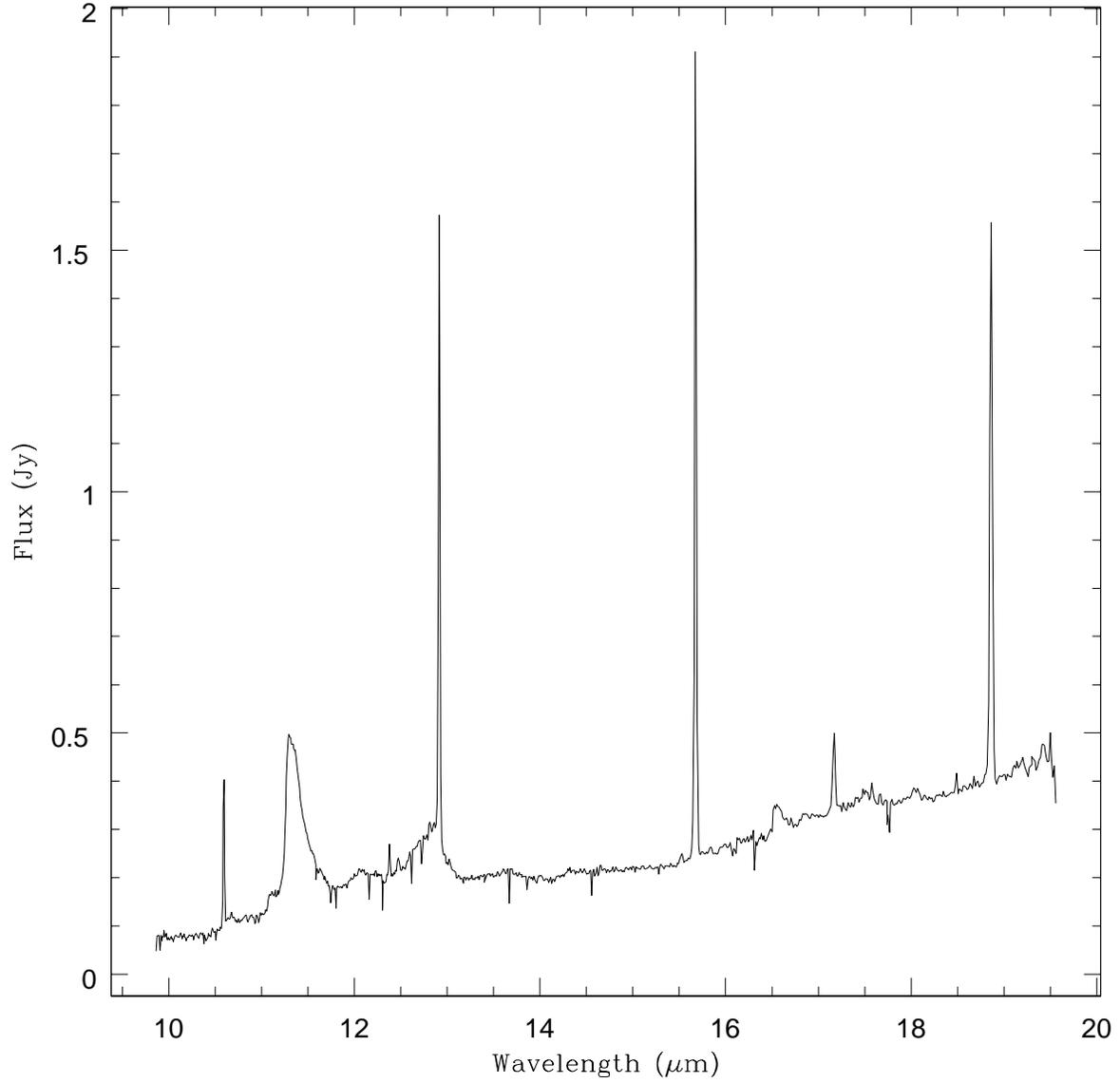}
\caption{The spectrum of NGC 1222 observed with IRS on SPITZER, reduced with SPICE. Order overlaps and noise spikes have been removed.}
\label{spectrum}
\end{figure}

\begin{figure}

\epsscale{0.50}
\plotone{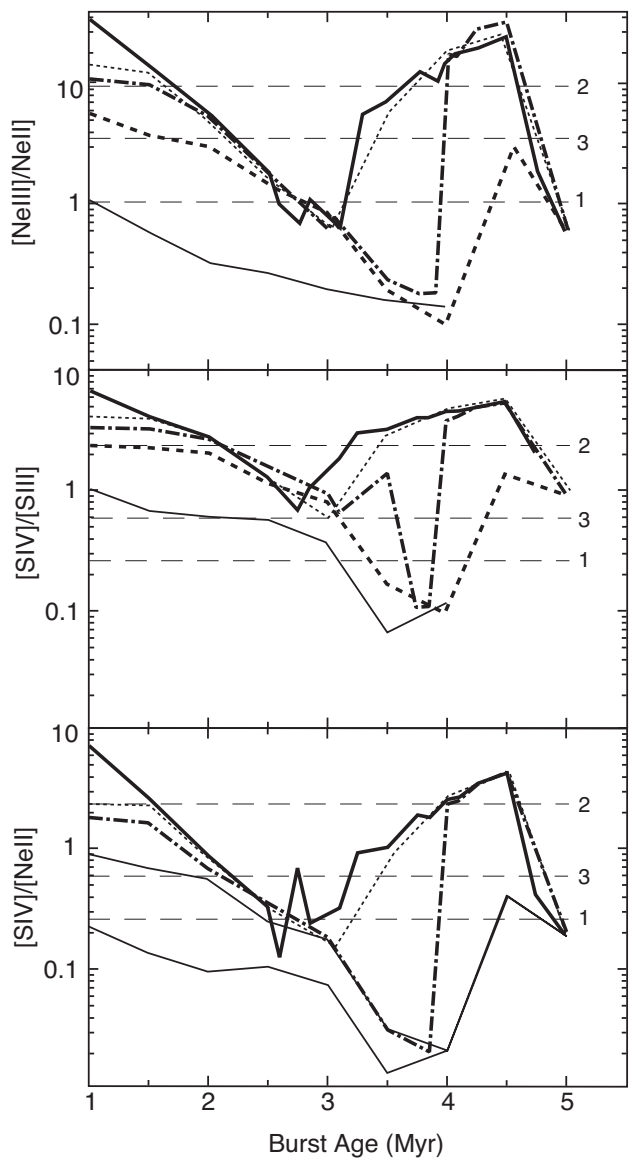}
\caption{Line ratios calculated from STARBURST99+MAPPINGS simulations with
observed line ratios.  From top to
bottom starting at the left
of each figure are the derived ratios as a function of time for starbursts with upper mass limits of
100, 70, 50, 40, and 30 $M_{\odot}$.  Horizontal lines are the ratios observed in NGC~1222
for (top to bottom) Model 2, Model 3, Model 1.}
\label{ratios}
\end{figure}

\begin{figure}
\epsscale{0.50}
\plotone{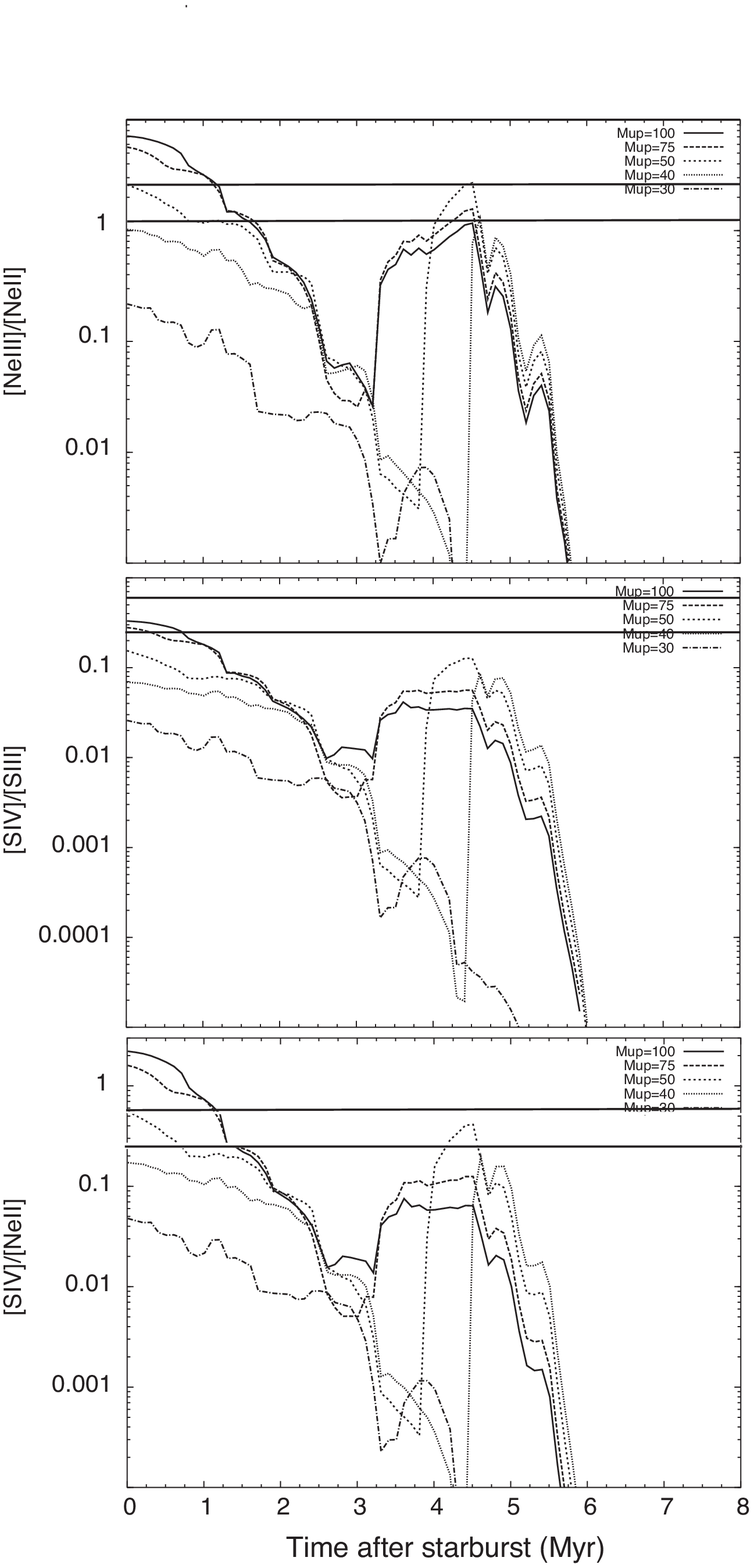}
\caption{Line ratios found with CLOUDY, from Rigby \& Rieke 2004. 
Horizontal lines are the observed NGC 1222 line ratios for Model 3 (top line)  and  
Model 1. Model 2 (the most concentrated) would in all cases be off the top of the graph.} 
\label{R&R}
\end{figure}

\clearpage
\begin{deluxetable}{lcccc}
\tabletypesize{\scriptsize}
\tablewidth{0pt}
\tablecaption{Radio Observations}
\tablehead{
\multicolumn{1}{l}{Band} &
\multicolumn{1}{l}{Program} &
\multicolumn{1}{c}{Date} &
\multicolumn{1}{c}{[Beam Size } &
\multicolumn{1}{c}{r.m.s.} 

\\
\colhead{Ghz}&
\colhead{}&\colhead{}&\colhead{$\arcsec$}&\colhead{mJy}

}
\startdata

 C, 5 &  AS286 & 11/23/87& ${1.75\times1.2}$ & 0.05 \\

K, 22 & AT309 & 6/24/05& ${0.96\times0.53}$ & 0.17\\

\enddata

\end{deluxetable}

\begin{deluxetable}{lcccc}
\tabletypesize{\scriptsize}
\tablewidth{0pt}
\tablecaption{Infrared Lines in NGC 1222}
\tablehead{
\multicolumn{1}{l}{Observation} &
\multicolumn{1}{c}{[NeIII] $15.55\mu$m} &
\multicolumn{1}{c}{[NeII] $12.81\mu$m} &
\multicolumn{1}{c}{[SIV] $10.51\mu$m} &
\multicolumn{1}{c}{[SIII] $18.71\mu$m}

\\
\colhead{}&
\colhead{Jy}&\colhead{Jy}&\colhead{Jy}
&\colhead{Jy}
}
\startdata 
9071872-0006-5-E1394294 &  1.7 & 1.35 & 0.32 & 1.1\\

9071872-0007-5-E1394295  & 1.72 & 1.6 & 0.38 & 1.37\\
\enddata

\end{deluxetable}

\begin{deluxetable}{lccc}
\tabletypesize{\scriptsize}
\tablewidth{0pt}
\tablecaption{Modeled Line Ratios: Compact Source}
\tablehead{
\multicolumn{1}{l}{Spatial Model} &
\multicolumn{1}{l}{[NeIII]/[NeII]} &
\multicolumn{1}{c}{[SIV]/[SIII]} &
\multicolumn{1}{c}{[SIV]/[NeII]} 

\\
}
\startdata
Model 1 \tablenotemark{a} & 1.17 & 0.28 & 0.24 \\
         
Model 2 \tablenotemark{b}  & 10.4  & 2.6  &    2.1   \\
             
Model 3 \tablenotemark{c}           &  2.6  & 0.63   &  0.53  \\
           
\enddata
\tablenotetext{a}{The galaxy is treated as one source}
\tablenotetext{b}{$90\%$ of [NeIII], [SIV] and $10\%$ of [NeII] and [SIII] in compact source.}
\tablenotetext{c}{$90\%$ of [NeIII], [SIV] and $40\%$ of [NeII] and [SIII] in compact source.}
\end{deluxetable}

\begin{deluxetable}{lccc}
\tabletypesize{\scriptsize}
\tablewidth{0pt}
\tablecaption{Line Ratios: Diffuse Emission}
\tablehead{
\multicolumn{1}{l}{Spatial Model} &
\multicolumn{1}{l}{[NeIII]/[NeII]} &
\multicolumn{1}{c}{[SIV]/[SIII]} &
\multicolumn{1}{c}{[SIV]/[NeII]} 

\\
}

\startdata
Model 1  & 1.13 & 0.25 & 0.26  \\
         
Model 2 & 0.14 & 0.02  &    0.026    \\
             
Model 3        &  0.19  & 0.04 &  0.04  \\
           
\enddata

\end{deluxetable}

\clearpage

\end{document}